\documentstyle[aps,epsf,multicol,eqsecnum]{revtex}
\begin{document}
\draft
\widetext
\title{Landau-Ginzburg Theories for Non-Abelian
Quantum Hall States}
\author{Eduardo Fradkin$^{1,4}$, Chetan Nayak$^{2,4}$,
and Kareljan Schoutens$^{3}$}
\address{
$^1$Department of Physics,
University of Illinois at Urbana-Champaign,
Urbana, IL 61801-3080, USA\footnote{Permanent Address}\\
$^2$Physics Department, University of California, Los Angeles
CA 90095-1547, USA$^*$\\
$^3$Institute for Theoretical Physics,
Valckenierstraat 65, 1018 XE Amsterdam, THE NETHERLANDS \\
$^4$Institute for Theoretical Physics, University of California,
Santa Barbara, CA 93106-4030
}
\maketitle
\begin{abstract}
We construct Landau-Ginzburg effective field theories
for fractional quantum Hall states -- such as the Pfaffian
state -- which exhibit non-Abelian statistics.
These theories rely on a Meissner construction
which increases the level of a non-Abelian
Chern-Simons theory while simultaneously
projecting out the unwanted degrees of
freedom of a concomitant enveloping Abelian theory.
We describe this construction in the context
of a system of bosons at Landau level filling
factor $\nu=1$, where the non-Abelian symmetry is a
dynamically-generated $SU(2)$
continuous extension of the discrete particle-hole
symmetry of the lowest Landau level.
We show how the physics of quasiparticles
and their non-Abelian statistics arises in
this Landau-Ginzburg theory. We describe its relation
to edge theories -- where a coset construction
plays the role of the Meissner
projection -- and discuss extensions to other
states.
\end{abstract}
\bigskip


\begin{multicols}{2}
\narrowtext
\section{INTRODUCTION}
\label{sec:intro}

It is possible for particles in $2+1$-dimensions
to have non-Abelian braiding statistics\cite{mr,wen}.
This means the following: there is not a unique
state, but rather a degenerate
set of states ${\Psi_A}$, describing a system with
$n$ particles at
${x_1}, {x_2}, \ldots, {x_n}$.
The result of exchanging or braiding
these particles need not be the mere
accrual of a phase
${\Psi_A} \rightarrow {e^{i\alpha}}{\Psi_A}$.
It will, in general, be the transformation
of these states within their degenerate subspace,
${\Psi_A} \rightarrow {M_{AB}}{\Psi_B}$.
Since the matrices, ${M_{AB}}$, corresponding to
different exchanges might not commute, we call
this non-Abelian statistics.

Clear indications of a quantum Hall plateau
at $\nu=5/2$ offer the tantalizing possibility
that non-Abelian statistics occurs in nature.
If this plateau is in the universality
class of the Pfaffian state \cite{mr} -- which is
one of the leading candidates -- then its
quasihole and quasiparticle excitations
will exhibit non-Abelian statistics,
as was shown in \cite{nw,fntw}. While these analyses
succeeded in establishing the non-Abelian statistics
and in making definite experimental predictions,
they left a number of unanswered questions.

In an earlier paper\cite{fntw}, a `dual' effective theory
of the Pfaffian state was constructed. However, the drawback of
the dual theory is that it takes the quasiparticles
as the fundamental objects. We seek a Landau-Ginzburg theory
since it would illuminate the unusual
physics of this state by showing how such objects
can arise from the underlying electron degrees of
freedom. Such a formulation would also facilitate
probes of the robustness of the state by
introducing perturbations which
couple in a local way to the electrons.

What makes this question so interesting is
precisely the fact that its answer is, at first glance, so
elusive. After all, electrons do not {\it a priori}
have any non-Abelian structure, so it is hard to see
from whence it could arise. (The approximate
non-Abelian symmetries -- such as spin or layer
symmetry -- of multi-component quantum Hall systems
are fundamentally Abelian insofar as their effect on
braiding statistics is concerned. See below.)

One piece of the puzzle is that the relevant non-Abelian
structures result from constraining -- or projecting
out part of -- an Abelian theory. This may be seen
most simply at the edge, where the neutral sector
of the theory, which contains the non-Abelian
statistics and has $c=1/2$, is `half' of
a $c=1$ Abelian theory. A second piece of the puzzle
is that a particularly simple and efficient
way of enforcing a constraint is through the
Meissner effect which, in the case of an ordinary
superconductor, imposes ${\bf B} = 0$. One salient feature of the ground states
that
have excitations with non-Abelian statistics is that they exhibit {\it
pairing}\cite{pairing}. Thus, at least naively, it is natural to expect that
these
states, generically called paired Hall states, should have Landau-Ginzburg
description
which should make the physics of pairing and of the Meissner effect manifest.
(Note that
the Meissner effect is already used to impose the basic charge-flux
commensuration.
What we propose is to also use the Meissner effect in the neutral sector to
project out
unwanted degrees of freedom.)

 On general grounds one expects that
the Landau-Ginzburg theory should contain a Chern-Simons term in its action for
the
gauge field associated with the relevant hydrodynamic current. It was shown by
Wen and
Zee\cite{wz} that for generic Abelian FQH states the effective action has the
form of a
multicomponent Abelian Chern-Simons action with symmetry $U(1)\times \ldots
\times
U(1)$. In some cases the system  is such that the symmetry can be promoted to a
larger
non-Abelian symmetry group, such as $SU(2)$, but at level $1$. In other words,
the coefficient of the non-Abelian Chern-Simons term, is determined by the
(level) index $k=1$. However, the state is still Abelian in the sense that the
all
representations of the Braid group at level $1$ associated with this state are
one-dimensional. It has been known since the seminal work of
Witten\cite{witten} on
Chern-Simons theory that only non-Abelian Chern-Simons gauge theories at level
$k \geq
2$ exhibit non-Abelian statistics. Thus, $SU(2)$ invariant FQH states such as
the
singlet FQH state\cite{bf,lf} at $\nu=1/2$, whose effective action contain an
$SU(2)$
invariant Chern-Simons term at level $1$, are actually Abelian from the point
of view
of the statistics. We will show in this paper that it is possible to use
the Meissner effect as a mechanism for generating an effective action at level
$k \geq
2$ for an unbroken hidden gauge symmetry. Thus, in this sense, pairing implies
non-Abelian statistics.

In \cite{ho}, it was pointed out that the
Pfaffian state
\begin{equation}
{\Psi_{\rm Pf}}\,\, =\,\,
{\rm Pf}\left(\frac{1}{{z_i} - {z_j}}\right)
\,\,\,{\prod_{i>j}}{\left({z_i} - {z_j}\right)^2}\,\,\,
{e^{-\frac{1}{4{\ell}_0^2} \sum |z_i|^2 }}
\label{Pfaffian}
\end{equation}
(where ${\rm Pf}\left(\frac{1}{{z_i} - {z_j}}\right) =
{\cal A}\left(\frac{1}{{z_1} - {z_2}}\,
\frac{1}{{z_3} - {z_4}}\,\ldots\right)$ is the antisymmetrized product
over pairs of electrons)
is closely related to an Abelian state, the $(3,3,1)$
state \cite{halperin},
\begin{eqnarray}
{\Psi_{(3,3,1)}}&=&
{\rm Pf}\left(\frac{{u_i}{v_j}+{v_i}{u_j}}{{z_i} - {z_j}}\right)
\,\,\,{\prod_{i>j}}{\left({z_i} - {z_j}\right)^2}\,\,\,
{e^{-\frac{1}{4{\ell}_0^2} \sum |z_i|^2 }}
\nonumber \\
&& \\
&=&{\cal A}\left\{
{\prod_{i>j}}{\left({z_{2i-1}} - {z_{2j-1}}\right)^3}\,
{\left({z_{2i}} - {z_{2j}}\right)^3}\,\,
\right.
\nonumber \\
&&
\left.
\times \; {\prod_{i,j}}{\left({z_{2i-1}} - {z_{2j}}\right)^1}\,
{\prod_i}{u_{2i-1}}{v_{2i}}\right\}
\,\,\,
{e^{-\frac{1}{4{\ell}_0^2} \sum |z_i|^2 }}
\nonumber \\
&&
\label{(3,3,1)}
\end{eqnarray}
where $u,v$ are up- and down-spin spinors and
the second equality follows from
the Cauchy identity.

The difference between the states can be
elucidated by quantizing the spins along
the $x$-axis
\begin{eqnarray}
\lefteqn{ {\Psi_{(3,3,1)}}\,\, =\,\, }
\nonumber \\
&& \quad
{\rm Pf}\left(\frac{{u^x_i}{u^x_j}-{v^x_i}{v^x_j}}{{z_i} - {z_j}}\right)
\,\,\,{\prod_{i>j}}{\left({z_i} - {z_j}\right)^2}\,\,\,
{e^{-\frac{1}{4{\ell}_0^2} \sum |z_i|^2 }}
\label{331bis}
\end{eqnarray}
Then, it is clear that the Pfaffian state
can be obtained from the $(3,3,1)$ state
by polarizing the (pseudo)spins along the
$x$-direction or, in other words, by projecting
out all of the down (via-a-vis the $x$-axis)
spins.

Unfortunately, it is not so clear how to
implement such a constraint using the Meissner effect,
since the Meissner effect is best suited for
a constraint such as $A=0$ which, in a Chern-Simons
theory, is equivalent to the condition that the associated
(hydrodynamic) current vanishes, $j=0$.
However, there is a bosonic analogue of (\ref{Pfaffian}) in which this
constraint assumes the desired form.
In [\ref{bib:fntw}], we took advantage of the fact that
a Pfaffian state of bosons at $\nu=1$,
\begin{equation}
{\Psi_{{\rm Pf},\nu=1}}\,\, =\,\,
{\rm Pf}\left(\frac{1}{{z_i} - {z_j}}\right)
\,\,\,{\prod_{i>j}}{\left({z_i} - {z_j}\right)}\,\,\,
{e^{-\frac{1}{4{\ell}_0^2} \sum |z_i|^2 }}
\label{bosePfaffian}
\end{equation}
has a dynamical $SU(2)$ symmetry which mixes the
charged and neutral sectors. The $SU(2)$ symmetry
is at level $k=2$ -- unlike the spin-rotational
$SU(2)$ symmetry of a spin-singlet state such
as the $(3,3,2)$ state which, since it has
$k=1$, can only lead to Abelian statistics.

However, Abelian states with a non-Abelian symmetry group are still useful in
the
following sense. In a number of cases of interest the dynamical
symmetry arises as follows. The charge current, which is the main observable in
the
Abelian classification of the FQH states, can be regarded as a diagonal
generator of a non-Abelian group (such as $SU(2)$). Thus an Abelian state with
symmetry
$U(1) \times U(1)$  can be described in terms of a system with an $SU(2)_1$,
even
though this symmetry in general is broken explicitly at the level of the
Hamiltonian.
An example of this construction is given in references [\ref{bib:bf}] and
[\ref{bib:lf}]
for the $(3,3,2)$ state.

In this paper we will discuss a different example in which the
$SU(2)$ symmetry is related to particle-hole symmetry. This is the case of the
(Abelian)
FQH
state for {\it bosons} at $\nu=1/2$. This is a bosonic
Laughlin state and it is the
bosonic analog of the $\nu=1/3$ state for (fully polarized) fermions.
Once this is
established we will show that the non-Abelian bosonic state at $\nu=1$, which
has as
its effective action an
$SU(2)_2$ Chern-Simons theory, can be constructed as the paired state resulting
from two
$\nu=1/2$ Abelian bosonic states,
with symmetry $SU(2)_1 \times SU(2)_1$.
We will see that pairing leads to a
Meissner effect in the broken sector of
 $SU(2)_1
\times SU(2)_1$ which
leads to an $SU(2)_2$ Chern-Simons theory in the unbroken
sector. We
will show that this mechanism of projection from an enveloping Abelian theory,
is the
counterpart in the bulk of the {\it conformal embedding}\cite{ce} of $SU(2)_2$
into
$SU(2)_1 \times SU(2)_1$ of the conformal field theory of the edge states. The
properties of this  bosonic theory can be worked out quite explicitly,
including the
properties of the quasiholes which are realized as solitons of the broken
symmetry
state. The generalization of these ideas to the physically more relevant (but
more
involved) fermionic state will be discussed elsewhere.

In essence, the desired Landau-Ginzburg theory should have the following
ingredients:
\begin{enumerate}
\item
{\it An $SU(2)_2$ Chern-Simons gauge field.}
\item
{\it Projection from an enveloping Abelian theory.} This
projection would naturally take the form of
projecting out the down spins in a two-component
Abelian state.
\item
{\it Pairing Physics.} There is a striking analogy (which we discuss below)
between the Pfaffian state and a p-wave BCS superconductor which explains, for
instance, the halving of the flux quantum.
\end{enumerate}
We will see below how these three ingredients can lead to a Landau-Ginzburg
theory for
the Pfaffian state.

Recently, the relation between pairing (and its generalizations), the
Pffafian states (and their generalizations) and Conformal Field Theory
has been reexamined by Read and Rezayi \cite{RR1,RR2}
who introduced quantum Hall states with $k$-particle condensates
(a similar idea was discussed earlier by Wen \cite{wen}), Wen and Zee
\cite{wz2},
and also by Cappelli, Georgiev and Todorov\cite{cappelli} who
discussed the relation between the
conformal counterparts at the edge of the Abelian $(3,3,1)$ state
and the Pfaffian.

This paper is organized as follows. In section \ref{sec:1/2} we show
that the Abelian Laughlin state of bosons at $\nu=1/2$ is described by
an $SU(2)$ Chern-Simons theory at level $1$.
Here we also discuss the effective theory for the $(2,2,0)$
bilayer bosonic state and show that it is an $SU(2)_1 \times SU(2)_1$
Chern-Simons theory. In section
\ref{sec:pair} we give the main arguments for the derivation of the
Landau-Ginzburg theory for the bosonic non-Abelian FQH states
and their relation with pairing. For simplicity
we will discuss the case of a symmetry group $SU(2)$ at level $2$ and we will
only state its generalizations. Here we show the connection between
pairing, the Meissner effect, the breaking of the symmetry $SU(2)_1 \times
SU(2)_1 \to
SU(2)_2$ and the physical origin of non-Abelian statistics. In section
\ref{sec:braid} we explore the consequences of the Landau-Ginzburg
theory and show that the quasihole excitations with
non-Abelian statistics are topological solitons of the Landau-Ginzburg theory.
In section
\ref{sec:edge} we discuss the bulk-edge connection for these states. In
particular we show that the projection procedure implicit in
the conformal embedding  of $SU(2)_2$ into $SU(2)_1 \times SU(2)_1$ at
the boundary is the counterpart of the Meissner effect in the bulk. We
also discuss generalizations of these results for multi-component
theories associated with the $q$-Pfaffian and Read-Rezayi states and their
edge states. Section \ref{sec:conc} is devoted to the conclusions.

\section{Bosons at $\nu=1/2$ and $SU(2)_1$}
\label{sec:1/2}

The ground state of a system of charged bosons (of charge $-e$) with strong
repulsive interactions in a large magnetic field exhibits the FQHE. For a
half-filled
lowest Landau level, $\nu=1/2$, the wave function of the ground state of this
system is
in the universality class of the bosonic Laughlin state
\begin{equation}
\Psi_{1/2}=\prod_{i<j}(z_i-z_j)^2 e^{-{\frac{1}{4\ell_0^2}}\sum_{i=1}^N
|z_i|^2}
\label{eq:Psibose1}
\end{equation}
The effective action of this FQH state is
\begin{equation}
S_{1/2}={\frac{2}{4\pi}}\int d^3x \;\;\;\epsilon_{\mu \nu \lambda} a^\mu
\partial^\nu
a^\lambda
\label{eq:S1/2}
\end{equation}
The conformal field theory of the edge states of this FQH state is a $c=1$
chiral
boson with compactification radius $R=1/\sqrt{2}$.

We want to show that this system has a hidden $SU(2)$ symmetry and that the
effective
action is that of a Chern-Simons theory for $SU(2)_1$. That such a
description must exist can be inferred from the structure of the Hilbert
space of the edge states at $\nu=1/2$. In fact, the edge states for this
FQH state of bosons can be described in terms of a  chiral Bose field $\phi$
with the
standard Lagrangian for chiral bosons,
\begin{equation}
{\cal L}_{\rm edge}={\frac{1}{4\pi}} \partial_x \phi \left(\partial_t
\phi-v \partial_x \phi \right)
\label{eq: Ledge}
\end{equation}
where $v$ is the velocity of the edge states (hereafter we will set $v=1$).

The edge charge density is
$\rho=e {\frac{\sqrt{\nu}}{2\pi}}  \partial_x \phi$, with $\nu=1/2$. The
operator that creates particle states with charge $e$ and Bose statistics is
\begin{equation}
\psi_{1}=e^{{\frac{i}{\sqrt{\nu}}}\phi}
\label{eq:p}
\end{equation}
Its presence implies that the compactification radius of the chiral boson
is $R=\sqrt{\nu}=1/\sqrt{2}$. In addition, the spectrum contains
quasiparticle states, created by the operator
\begin{equation}
\psi_{1/2}=e^{i{\sqrt{\nu}}\phi}
\label{eq:qp}
\end{equation}
which creates states with charge $e/2$ and statistics $\pi/2$ (semions).

It is a well known fact\cite{ginsparg} that, precisely at $\nu=1/2$, the
charge density
operator $\rho$, the particle (charge $e$) operator $\psi_{1}$ and its adjoint
$\psi_{1}^\dagger$ have (the same) scaling dimension $1$, and span the
triplet $S=1$ representation of $SU(2)$. From the point of view of
the current algebra, this state supports an $SU(2)$ chiral Kac-Moody algebra
at level $1$ with $J^\pm = e^{\pm i \sqrt{2} \phi}$ and $J^3
={\frac{1}{\sqrt{2}}} \partial_x \phi$.
Notice the fact, important for what follows, that the charge density (or
current, since
this is a chiral theory) is the diagonal (Cartan) generator of $SU(2)$.

The conformal field theory associated with the $SU(2)_1$ current algebra is the
chiral Wess-Zumino-Witten model\cite{witten-bosonization} (at level $k=1$) with
action
\begin{eqnarray}
S_{WZW}=&&{\frac{k}{16 \pi}} \int_{S^2} d^2x \; {\rm Tr}\; \partial_\mu g
\partial^\mu g^{-1}
\nonumber \\
+&&{\frac{k}{24\pi}} \int_B d^3y \; \epsilon^{ijk} {\rm Tr}\;
{\bar g}^{-1} \partial_i {\bar g} \;
{\bar g}^{-1} \partial_j {\bar g} \;
{\bar g}^{-1} \partial_k {\bar g}
\nonumber \\
&&
\label{eq:wzw}
\end{eqnarray}
where $g$ is an $SU(2)$-valued chiral field defined on the sphere $S^2$, ${\bar
g}$ is its extension from the sphere to the ball $B$, and $k \in {\bf Z}$
is the level.

On the other hand, from Witten's work on Chern-Simons
theory\cite{witten}, we also know that the chiral Wess-Zumino-Witten model (at
level $k$)
is equivalent to the $SU(2)$ Chern-Simons gauge theory on a disk (also at level
$k$).
Hence, from this point of view, it should be possible to map the physics
of the {\it bulk} state of $\nu=1/2$ bosons onto an $SU(2)_1$
Chern-Simons gauge theory whose action is given by
\begin{eqnarray}
S_{\rm CS}(a)=&&\frac{1}{4\pi} \,\int\,{d^3}x\,{\epsilon^{\mu\nu\lambda}}\,
\left({a_{\mu}^a}{\partial_\nu}{a_{\lambda}^a} + \frac{2}{3}
{f_{abc}}{a_{\mu}^a}{a_{\nu}^b}{a_{\lambda}^c}\right) \nonumber \\
&&
\label{eq:nacs}
\end{eqnarray}
Since the diagonal generator of $SU(2)$ is the charge
current, a coupling to an external electromagnetic field breaks
explicitly this $SU(2)$ symmetry.
In this language, the quasiparticle of the bulk $\nu=1/2$ state is
created by the Wilson loop operator in the fundamental ($S=1/2$) representation
of
$SU(2)$. This state carries charge $e/2$ and $\pi/2$ statistics and it is a
one-dimensional representation of the Braid group. At level $k=1$, all the
representations of the Braid group are Abelian\cite{witten}.

Hence, we conclude that the effective low-energy theory of the
$\nu=1/2$ FQH effect
for bosons is the $SU(2)_1$ Chern-Simons gauge theory of Eq.\ (\ref{eq:nacs}).
The mapping we have just discussed indicates that it should be possible to
construct the $\nu=1/2$ bosonic FQH state in terms of an $SU(2)_1$
Chern-Simons gauge theory.
In order to do that, we first notice that {\it hard core}
bosons at half-filling are invariant under particle-hole symmetry.
To make this fact manifest we introduce the boson creation and
annihilation operators, $B^\dagger$ and
$B$, and the boson number operator $B^3$ in terms of the doublet of boson
operators
$\psi_1$ and $\psi_2$ , such that
\begin{eqnarray}
B^\dagger&=&\psi_1^\dagger \psi_2 \nonumber \\
B&=&\psi_2^\dagger \psi_1 \nonumber \\
B^3&=&\psi_1^\dagger \psi_1-\psi_2^\dagger \psi_2 \nonumber \\
&&
\end{eqnarray}
which is a representation of the $SU(2)$ algebra. In addition we impose the
constraint
\begin{equation}
\psi_1^\dagger \psi_1 +\psi_2^\dagger \psi_2=1
\label{eq:single}
\end{equation}
which is the ``hard-core" condition. Notice that, in this representation, the
diagonal
generator of $SU(2)$ is identified with the electric charge. With this choice,
the coupling to an external electromagnetic field in general  breaks this
$SU(2)$
(particle-hole) symmetry since it couples only to the diagonal generator.
But, {\it at half-filling} the average external magnetic field is exactly
canceled
by the expectation value of the $SU(2)$ gauge field (induced by the
bosons). Hence, at half-filling $SU(2)$ invariance is exact. In particular, the
half-filling condition reads
\begin{equation}
\int B_3=0
\label{eq:neutrality}
\end{equation}
which in this notation is just a neutrality condition.
Notice also that $SU(2)$ transformations rotate bosons of type 1 into bosons of
type 2
and vice versa. Given the constraint Eq.\ (\ref{eq:single}) these are
particle and hole states. This structure is essential to the
construction that we will pursue here.

The neutrality condition Eq.\ (\ref{eq:neutrality}) ({\it i.\ e.\/}
half-filling)
implies that the ground state of the $SU(2)$ doublet bose field $\psi$ is
strictly
charge conjugation invariant. Thus, although microscopically the $SU(2)$
triplet
(adjoint) Bose field $B$ is a
non-relativistic field, the effective Lagrangian for the low-energy
physics associated with this state must be such that it respects charge
conjugation invariance. In particular this means that this Bose field is
not in a Bose condensed state. Hence, the excitations of
this state are the bosons themselves. In this notation, the Laughlin
state $|\Psi_{1/2}\rangle$ reads
\begin{equation}
|\Psi_{1/2}\rangle= \prod_{i<j} \; (z_i-z_j)^2 \; \prod_i {B^\dagger}({z_i})
|0\rangle
\label{eq:psi1/2}
\end{equation}
The effective generalized Haldane Hamiltonians for (hard-core) bosons in
the lowest Landau level are closely related to the integrable
one-dimensional spin-$1/2$ chains and the Haldane-Shastry model. In
fact, the wave function for the $\nu=1/2$ FQH state of bosons can be
thought of as the wave function for a ``spin" system where the complex
numbers $z_i$ are the coordinates of, say, the down
spins\cite{haldane-analogy}. In this picture, the charge becomes the $z$
component of the spin. Using Haldane's method of
pseudo-potentials\cite{pseudo},
it is possible to construct a local Hamiltonian for hard-core
bosons at $\nu=1/2$ which has the topological fluid state
$|\Psi_{1/2}\rangle$ as its ground state.
The spectrum of these Hamiltonians, which are fully gapped, is also
generated by an $SU(2)_1$ Kac-Moody current algebra much as in the case of
the conventional (gapless) one-dimensional quantum Heisenberg antiferromagnets
\cite{affleck}.
Microscopically, realistic Hamiltonians for the bulk system
do not in general have a
continuous $SU(2)$ symmetry (a discrete symmetry is sufficient to
satisfy the requirement of particle-hole symmetry). However,
the ground state $\Psi_{1/2}$ is actually an $SU(2)$ singlet
and the full $SU(2)$ symmetry is not spontaneously broken.
However, unlike the conventional $S=1/2$ (non-chiral)
quantum Heisenberg antiferromagnet, the spectrum of these chiral models
is fully gapped and, up to topological degeneracies, the ground state of
this topological fluid is unique. Thus, even if the microscopic
Hamiltonian is not fully $SU(2)$ invariant, the symmetry breaking
terms only modify the  energies of the bulk excitations of a given multiplet
but not their quantum numbers. At the edge, the $SU(2)$ symmetry is
exact.

The identification of the charge current with the diagonal generator of
$SU(2)$ and the  $SU(2)$ invariance of the low energy physics, combined with
the fact that the at the edge the boundary conformal field theory is an $SU(2)$
Wess-Zumino-Witten theory at level $1$,
require that the effective (Landau-Ginzburg)
Lagrangian for the $\nu=1/2$ FQH state for
bosons should contain an $SU(2)$ Chern-Simons
term at level 1.

Furthermore, the $\Psi_{1/2}$ state supports {\it
spinon} excitations which are {\it semions} that carry spin-$1/2$ ({\it i.\
e.\/} charge $e/2$). In the spin system picture, the quasiparticles are
{\it topological solitons} with semion statistics.
These states are the quasiparticles of the $\nu=1/2$ FQH state for (hard-core)
bosons.

$B^\dagger$ and $B$
are the boson creation and annihilation operators,
which are composites of quasiparticles of the same
sign, i.e. charge $-e/2$ and $e/2$ respectively.
$B_3$ creates a magnetophonon, which is a neutral composite
of quasiparticles of opposite sign. There are microscopic
models for which these three excitations are degenerate,
although discrete particle-hole symmetry only guarantees
the degeneracy between the two charged states. At the edge,
however, the $SU(2)$ symmetry and the resulting degeneracy
is always exact. Similarly, the quantum numbers -- principally
the statistics -- in the bulk also exhibit this symmetry
even when the energies do not.

The above considerations imply that Landau-Ginzburg effective
Lagrangian for the $\Psi_{1/2}$ state should have the form
\begin{eqnarray}
{\cal L}=&&|D B|^2+ |D b|^2 +
V(b)+ V(B)
\nonumber \\
&&
+1 \cdot {\cal L}_{\rm CS}(a)
+\epsilon^{\mu \nu \lambda} A_\mu f_{\nu \lambda}^3
\nonumber \\
&&
\label{eq:L1/2}
\end{eqnarray}
where $B$ is the $SU(2)$ triplet Bose field introduced above,
$b$ is a doublet field which carries the spinor representation of
$SU(2)_1$, $a_\mu$ is an $SU(2)$ gauge field, $D B$ is the covariant
derivative in the triplet (adjoint) representation of $SU(2)$, $D b$
is the covariant  derivative in the spinor (fundamental) representation,
$A_\mu$ is the electromagnetic perturbation and $f^3$ is the component
of the field tensor of the gauge field $a_\mu$ along the diagonal generator of
$SU(2)$. Finally, $V(b)$ and $V(B)$ are potentials that give masses,
and interactions, to the doublet and triplet excitations. These fields do
not condense and, in this phase, the $SU(2)$ gauge symmetry is exact.
We are also assuming here that the triplet excitations, which in a sense
are bound states of the doublet quasiparticle states, are stable. As we
stated above, this can be achieved in physical microscopic systems. In
what follows we will only indicate one matter field and, unless stated
otherwise, we will only discuss the triplet field since,
as we will show below, it is the one relevant to the Pfaffian state.

The form of the effective Lagrangian of Eq.\ (\ref{eq:L1/2})
is analogous to the Landau-Ginzburg
Lagrangian for the Laughlin states except that the gauge field
takes values on the $SU(2)$ algebra instead of
$u(1)$. In eq.\ (\ref{eq:L1/2}), ${\cal L}_{\rm CS}(a)$ is the Chern-Simons
action for the $SU(2)$ gauge field of Eq.\ (\ref{eq:nacs}).
Unlike its more familiar $U(1)$ relative\cite{zhk}, this effective
Landau-Ginzburg
theory does not break the $SU(2)$ symmetry.
The ``relativistic" form of the Lagrangian of Eq.\ (\ref{eq:L1/2})
(namely, a Lagrangian which is second order in time derivatives)
is a consequence of the particle-hole symmetry of this state.
In fact, this is the Lagrangian with the smallest possible number of
derivatives compatible with all the symmetries. The
potential is chosen in such a way that the ground state has $B_a=0$,
{\it i.\ e.\/} the vacuum is an $SU(2)$ singlet.

Hence, the excitations described by this effective theory have exactly
the same quantum numbers as the excitations of the (equivalent) Abelian
theory of Eq.\ (\ref{eq:S1/2}). In particular, the quasiparticles are
not bosons of charge $-e$ but {\it semions} with charge $-e/2$. By
particle-hole symmetry, there is also a hole with the opposite charge
and the same statistics. Thus, we have succeeded in showing that the
particle-hole symmetric Laughlin state for bosons at $\nu=1/2$ is
described by an effective Landau-Ginzburg theory with with a $SU(2)_1$
Chern-Simons term. In what follows we will make extensive use of this
result.

In the following section, we will construct the
bosonic pfaffian state Eq.\ (\ref{bosePfaffian}) by projecting out
unwanted degrees of freedom in the $(2,2,0)$ state, which is
a state of a bi-layer system in which the bosons
in the two layers simply form two independent
$\nu=1/2$ bosonic Laughlin states:
\begin{eqnarray}
{\Psi_{(2,2,0)}} &=&
{\rm Pf}\left(\frac{{u_i}{v_j}+{v_i}{u_j}}{{z_i} - {z_j}}\right)
{\prod_{i>j}}{\left({z_i} - {z_j}\right)}
{e^{-\frac{1}{4{\ell}_0^2} \sum |z_i|^2 }}\cr
&=&\,\,{\cal A}\left\{
{\prod_{i>j}}{\left({z_{2i-1}} - {z_{2j-1}}\right)^2}\,
{\left({z_{2i}} - {z_{2j}}\right)^2}\,\,
\right.
\nonumber \\
&& \;\;\;\;\;\;\;\;\;\;
\left. \times
{\prod_i}{u_{2i-1}}{v_{2i}}\right\}
\,\,\,
{e^{-\frac{1}{4{\ell}_0^2} \sum |z_i|^2 }}
\nonumber \\
&&
\label{(2,2,0)}
\end{eqnarray}
The relationship of this state to the bosonic Pfaffian
at $\nu=1$ is the same as that of the $(3,3,1)$
state to the fermionic Pfaffian at $\nu=1/2$:
it is the enveloping Abelian state. The Landau-Ginzburg
theory for this state is simply two copies
of Eq.\ (\ref{eq:L1/2}) with two bosonic fields
$B_1^i$, $B_2^i$, and its Lagrangian is
\begin{eqnarray}
{\cal L} &=&
 | \left(\partial + i{a_1} \right) {B_1}|^2 +\,
| \left( \partial + i{a_2} \right) {B_2}|^2 +
V(B_1)+V(B_2)
\nonumber\\
&&+\,\,\frac{1}{4\pi} {\epsilon^{\mu\nu\lambda}}\,
\left({a_{1\mu}^a}{\partial_\nu}{a_{1\lambda}^a} + \frac{2}{3}
{f_{abc}}{a_{1\mu}^a}{a_{1\nu}^b}{a_{1\lambda}^c}\right)\,
\nonumber \\
&&+
\frac{1}{4\pi} {\epsilon^{\mu\nu\lambda}}\,
\left({a_{2\mu}^a}{\partial_\nu}{a_{2\lambda}^a} + \frac{2}{3}
{f_{abc}}{a_{2\mu}^a}{a_{2\nu}^b}{a_{2\lambda}^c}\right)
\nonumber \\
&&+\,\frac{1}{2\pi}
{A_\mu}{\epsilon^{\mu\nu\lambda}} \left({f_1^{\nu\lambda 3}}
+ {f_2^{\nu\lambda 3}} \right)
\nonumber \\
&&
\label{eq:S}
\end{eqnarray}
In the next section we will show that the Pfaffian state has an
effective action which is derived from the effective action of Eq.\
(\ref{eq:S}) by a {\it pairing} of the (triplet) bosons $B_1$ and $B_2$.

\section{Pairing in the Quantum Hall Effect and
Landau-Ginzburg theory of the Pfaffian State of
Bosons at $\nu=1$ }
\label{sec:pair}

The Pfaffian state,
\begin{equation}
{\Psi_{\rm Pf}}\,\, =\,\,
{\rm Pf}\left(\frac{1}{{z_i} - {z_j}}\right)
\,\,\,{\prod_{i>j}}{\left({z_i} - {z_j}\right)^2}\,\,\,
{e^{-\frac{1}{4{\ell}_0^2} \sum |z_i|^2 }}
\end{equation}
and its generalizations,
\begin{equation}
{\Psi_{\rm qPf}}\,\, =\,\,
{\rm Pf}\left(\frac{1}{{z_i} - {z_j}}\right)
\,\,\,{\prod_{i>j}}{\left({z_i} - {z_j}\right)^q}\,\,\,
{e^{-\frac{1}{4{\ell}_0^2} \sum |z_i|^2 }}
\end{equation}
which we call the $q$-Pfaffian states, with $\nu={\frac{1}{q}}$
and $q$ even or odd for fermions or bosons, respectively,
are reminiscent of the real-space form of
the BCS wave function
\begin{equation}
{\Psi_{\rm BCS}} = {\cal A}\left\{g({{\bf r}_1}-{{\bf r}_2})
g({{\bf r}_3}-{{\bf r}_4}) \ldots \right\}
\end{equation}
This analogy was discussed at length
by Greiter, Wen, and Wilczek \cite{pairing}.
They pointed out that the Pfaffian state
can be thought of as the quantum Hall incarnation
of a superconductor with $p+ip$ pairing symmetry
since the pair wave function, $g({{\bf r}_i}-{{\bf r}_j})$,
is given by $1/({z_i} - {z_j})$.

As in a superconductor, the basic vortex carries
half of a flux quantum. A state with two such
half flux quantum quasiholes at $\eta_1$ and $\eta_2$
takes the form:
\begin{eqnarray}
\label{twoqhPfaffian}
{\Psi_{\rm qPf}}\,\, =\,\,
{\rm Pf}\left(\frac{\left({z_i}-{\eta_1}\right)\left({z_j}-{\eta_2}\right)
+ i\leftrightarrow j}
{{z_i} - {z_j}}\right)\cr
\,\,\,{ \times \prod_{i>j}}{\left({z_i} - {z_j}\right)^q}\,\,\,
{e^{-\frac{1}{4{\ell}_0^2} \sum |z_i|^2 }}
\end{eqnarray}

States with four or more half flux quantum
quasiholes exhibit non-Abelian statistics
\cite{nw,fntw}. One of the goals of this paper
is to show how these excitations and their non-Abelian
statistics arise naturally in a Landau-Ginzburg theory.
The above analogy between the Pfaffian
state and a $p+ip$ superconductor
guides us to look for a Landau-Ginzburg theory
of a paired order parameter. We shall here
concentrate on the case $q=1$, which is the
bosonic Pfaffian at $\nu=1$.

In the Pfaffian state, it is not the bosons $B_1$, $B_2$
which condense, but rather, we expect, a paired
order parameter such as $\Psi_a$:
\begin{equation}
{\Psi_a}\left({z_1}-{z_2}\right) =
{\epsilon_{abc}}{B_1^b}\left({z_1}\right)
{B_2^c}\left({z_2}\right) \,
f\left({z_1}-{z_2}\right)
\end{equation}
Here, $f$ is a maximally $T$-violating
$p$-wave pairing kernel which falls
off as $1/({z_1}-{z_2})$ at long distances.
$\Psi_a$ transforms in the spin-$1$ representation
of both $SU(2)$'s and in the spin-$1$ representation
of the diagonal $SU(2)$ subgroup.

We can equally use a non-linear realization of
the symmetry:
\begin{equation}
O = {e^{i{T_a}{\Psi_a}}}
\end{equation}
where the $T_a$'s are the $SU(2)$
generators in the spin-$1$ representation.
$O$ transforms as:
\begin{equation}
O \rightarrow  {G_1}\, O {G_2^{-1}}
\end{equation}

The desired Landau-Ginzburg theory
can be derived from the theory of two
independent bosonic $\nu=1/2$'s by coupling
them with an effective interaction of the form;
\begin{equation}
-\int d^3x \; g |{\vec B}_1 \times {\vec B}_2|^2
\end{equation}
where g is a coupling constant.
Next we introduce $O$ as a Hubbard-Stratonovich
field to decouple this quartic interaction term.
As a result, the order parameter field $\Psi$ picks up
an expectation value and the Hubbard-Stratonovich field $O$ factors into
an amplitude field (determined by the expectation value of the order parameter)
and a field which is an element of $SU(2)$ in the adjoint representation,
which hereafter we will call $O$. A gauge-invariant kinetic term for $O$
is generated by integrating out the high-energy modes of $B_1$, $B_2$.
As a result we have a Lagrangian of the form:
\begin{equation}
\int d^3x \left\{ \kappa {\rm Tr}\left({O^{-1}} {{D_\mu}O}\,
{O^{-1}}{{D_\mu}O}\right)\,+
\lambda {B_1}\,O\,{B_2} \right\}
\label{eq:coupling}
\end{equation}
where $\kappa$ is the rigidity of this broken symmetry state and the
coupling constant $\lambda$ is a (smooth) function of the coupling
constant $g$ and of the expectation value of the amplitude of the
order parameter. In the spirit of a non-linear sigma model for the low energy
physics, we will absorb this expectation value in the rigidity $\kappa$ and
in the coupling constant $\lambda$, and will ignore all amplitude
fluctuations.
In Eq.\ (\ref{eq:coupling}) $D_\mu$ is the covariant derivative,
${D_\mu}={\partial_\mu}+i ({a_{1\mu}}-{a_{2\mu}})$, which reflects the
way  the order parameter field $O$ couples to $SU(2) \times SU(2)$.

Hence, we see that when $O$ acquires an expectation
value, it breaks the $SU(2)\times SU(2)$
symmetry to its diagonal subgroup. As a result, the desired Meissner
constraint follows
and the combination of gauge fields $a^i_{1\mu}-a^i_{2\mu}$ acquires a
mass and its fluctuations decouple from the spectrum. In this low-energy limit,
the Meissner effect  on the diagonal symmetry implies a constraint of
the form
\begin{equation}
{a_{\mu}^i}\equiv {a_{1\mu}^i}={a_{2\mu}^i}
\end{equation}
then, by enforcing this constraint, we effect:
\begin{equation}
{{\cal L}_{CS}}({a_{1\mu}^i}) + {{\cal L}_{CS}}({a_{2\mu}^i})
\rightarrow \,2\,{{\cal L}_{CS}}({a_{\mu}^i})
\end{equation}
This is naturally accomplished by a Meissner construction
with symmetry-breaking pattern: ${SU(2)_1} \times
{SU(2)_1} \rightarrow {SU(2)_2}$. The simplest model which
incorporates this physics is of the form:
\begin{eqnarray}
S &=&
\int d^3x \left\{ \kappa {\rm Tr}\left({O^{-1}} {{D_\mu}O}\,
{O^{-1}}{{D_\mu}O}\right)\,+
\lambda {B_1}\,O\,{B_2} \right\}
\nonumber \\
&&+ \int d^3x  \left( | \left(\partial + i{a_1}\right) {B_1}|^2\,+
 | \left(\partial + i{a_2}\right) {B_2}|^2\right)
\nonumber \\
&&+ \,\,\frac{1}{4\pi} \,\int\,{d^3}x\,{\epsilon^{\mu\nu\lambda}}\,
\left({a_{1\mu}^a}{\partial_\nu}{a_{1\lambda}^a} + \frac{2}{3}
{f_{abc}}{a_{1\mu}^a}{a_{1\nu}^b}{a_{1\lambda}^c}\right)
\nonumber \\
&&+
\frac{1}{4\pi} \,\int\,{d^3}x\,{\epsilon^{\mu\nu\lambda}}\,
\left({a_{2\mu}^a}{\partial_\nu}{a_{2\lambda}^a} + \frac{2}{3}
{f_{abc}}{a_{2\mu}^a}{a_{2\nu}^b}{a_{2\lambda}^c}\right)
\nonumber \\
&&
\label{lg}
\end{eqnarray}
The matrix field $O$ is in the adjoint representation of $SU(2)$ and, as
such, it is blind to the center ${\bf Z}_2$ of $SU(2)$. Hence,
effectively $O\in SO(3)$, and it transforms under $SO(3)\times SO(3)$
according to
\begin{equation}
O \rightarrow {G_1}\,O\,{G_2^{-1}}
\end{equation}
{}From now on, whenever we discuss the field $O$ we will refer only to its
properties in $SO(3)$.

When $O$ acquires an expectation value, $\langle O\rangle = I$,
the symmetry is broken to the diagonal subgroup,
$SO(3)\times SO(3) \rightarrow SO(3)$. There is only
a single gauge field left, $a \equiv {\frac{1}{2}}({a_1} + {a_2})$,
corresponding to the unbroken symmetry group.
The Meissner effect sets the other combination,
${a_1} - {a_2}$ to zero. As a result, the gauge field
$a$ is promoted to level $2$ and it is responsible for the exotic braiding
statistics.

The other effect of the symmetry-breaking
is the polarization of the pseudo-spins
along the $x$-axis -- i.e. projecting out the
`down' spins in the bosonic analog of Eq.\
(\ref{331bis}). In the symmetry-broken state,
the second term in Eq.\ (\ref{lg}) is
${B_1}{B_2}$, which is just $\sigma_x$
(remember that we are referring to the
pseudo-spins, i.e. the layer index, {\it not}
the particle-hole $SU(2)$ degree of freedom).
Hence, the same symmetry-breaking which
gives us a level-$2$ gauge field also
projects out the unwanted degrees of freedom of
the enveloping Abelian theory.

So far we have assumed that the only possible phase transition out of
the Abelian $(2,2,0)$ state is through the mechanism of
the (triplet) pairing of bosons we just presented. Conceptually
there is however another possibility. Instad of a triplet paired state
of charge $e$ bosons, we can consider instead the {\it singlet} pairing
of the spin-$1/2$ semion {\it quasiparticles}. These quasiparticles are
described by doublet fields $b_\alpha$ ($\alpha=\pm1/2$),
as discussed in section \ref{sec:1/2}. The state that results is
a condensate of pairs of quasiparticles and, in a loose sense, it is
a hierarchical state\cite{ho}. We can repeat for this
state the construction that we presented in this section
for the triplet paired state, but now for a singlet paired field
$\Psi \propto b^\dagger_1 b_2$. This condensate also breaks
$SU(2)_1 \times SU(2)_1 \to SU(2)_2$. Thus, superficially this
may seem to
be another candidate for the Landau-Ginzburg theory for the Pfaffian
state. However, because of the coset $SU(2) \times SU(2)/SU(2)$ is
topologically trivial (see section \ref{sec:braid}), the spectrum of the
system in
this paired state does not contain the spin-$1/2$ representation which
are quasiparticles with non-Abelian
statistics. Hence, this is actually an Abelian paired state.

We conclude this section by generalizing the
construction that we just presented, to the case
in which there is more than one condensate. Specifically, we would
like to find a mechanism for the symmetry breaking pattern $SU(2)_1 \times
\ldots \times SU(2)_1 \to SU(2)_k$ ($k$ $SU(2)_1$'s to one $SU(2)_k$).
Thus we consider a system with ``$k$ layers" with a Lagrangian which is
the obvious generalization of Eq.\ (\ref{eq:S}) which includes
$k$ triplet fields $B_I$ ($I=1,\ldots,k$).  Now we consider the
situation in which these particles condense in overlapping pairs with
the pattern $B_1$ pairs with $B_2$, $B_2$ pairs with $B_3, \ldots ,
B_{k-1}$ pairs with $B_k$. Hence, all the $SU(2)$ symmetries are broken
except for the {\it diagonal} $SU(2)$ symmetry. Then, the same arguments
given in this section show that the symmetry becomes $SU(2)_k$.
The filling fraction of these states is $\nu={\frac{k}{2}}$. This
is the effective action of a generalization of the Pfaffian state that
was considered recently by Read and Rezayi\cite{RR2}.
In section \ref{sec:edge} we discuss the boundary conformal field theory
of this state and further generalizations.

\section{Braiding Statistics of Excitations in the Pfaffian State}
\label{sec:braid}

In the Chern-Simons Landau-Ginzburg theories of
Abelian quantum Hall states, the basic quasihole/quasiparticle
excitations manifest themselves as vortex solutions. The
same is true for the half flux quantum excitations
in the non-Abelian case, but the topological structure
is more complicated. In general, when a symmetry
group $G$ is spontaneously broken down to a residual
symmetry group $H$, there are topologically stable
vortex solutions which are classified by
the homotopy group ${\pi_1}(G/H)$ \cite{homotopy}.
In the $U(1)$ case, this is simply ${\pi_1}(U(1))={\bf Z}$;
the vortices are classified by winding number.
In the case of the Landau-Ginzburg theory for
the Pfaffian state, however, this is
${\pi_1}(SO(3)\times SO(3)/SO(3))$.
This homotopy group can be computed from the
long exact sequence
\begin{eqnarray}
\lefteqn{
\ldots \rightarrow {\pi_1}\left(SO(3)\right)
\rightarrow {\pi_1}\left(SO(3) \times SO(3)\right)}
\nonumber \\
&&
\rightarrow {\pi_1}\left(\left(SO(3) \times SO(3)\right) 
/ SO(3)\right)
\rightarrow {\pi_0}\left(SO(3)\right)\rightarrow \dots
\end{eqnarray}
associated to the fibration
\begin{eqnarray}
\lefteqn{
SO(3) \rightarrow SO(3) \times SO(3)}
\nonumber \\
&&
\qquad\qquad
\rightarrow \left(SO(3) \times SO(3)\right) / SO(3)
\end{eqnarray}
The result is
\begin{equation}
{\pi_1}(SO(3) \times SO(3)/SO(3)) = {\bf Z}_2
\end{equation}
Hence, there is one topologically distinct class
of vortex solutions of the Landau-Ginzburg theory
Eq.\ (\ref{lg}). These solutions have the asymptotic form:
\begin{equation}
O(r\rightarrow\infty,\theta) = {\cal R}({\hat {\bf n}},\theta)
\end{equation}
The gauge field, ${a_1}-{a_2}$ must follow $O$ in order for the
energy to be finite:
\begin{equation}
{a_{1\theta}} (r\rightarrow\infty,\theta) -
{a_{2\theta}} (r\rightarrow\infty,\theta) =
{{\cal R}^{-1}}({\hat {\bf n}},\theta)\,{\partial_\theta}
{\cal R}({\hat{ \bf n}},\theta)
\end{equation}

There is a different stable vortex solution
for any three-dimensional
unit vector, ${\hat {\bf n}}$; these solutions
are transformed into each other by rotations
in the {\it unbroken} $SO(3)$ subgroup.
At the quantum-mechanical level, this simply means
that there is an $SO(3)$ multiplet of  vortices,
{\it i.\ e.\/}  the vortices carry an $SO(3)$ quantum number.

The $SO(3)$ quantum number may be calculated
by the technique of Goldstone and Wilczek \cite{goldstone}.
We find that these vortices carry spin-1/2. Hence, they
exhibit the non-Abelian statistics discussed
in \cite{nw,fntw}.

In addition to these topological solitons, the
Landau-Ginzburg theory also contains the
fundamental fields, ${B^3}$, $B$ and $B^\dagger$.
These are now {\it fermionic} because they are
an $SU(2)$ triplet
coupled to a level $k=2$ $SU(2)$ gauge
field. The statistical transmutation
may be seen, following \cite{fntw},
by applying the results of \cite{witten}.
${B^3}$ is the neutral fermionic excitation
which we expect in a paired state. It can be visualized
as a fermionic dipole, which is the Pfaffian incarnation
of the fermionic dipoles discussed in the context of
compressible states\cite{dipole}.
${B}$ and $B^\dagger$ are the Pfaffian analogues of the
Laughlin quasiparticle
and quasihole. They carry a single flux quantum of
the $U(1)_c$ subgroup of $SO(3)$.
Together, ${B^3}$, $B$ and $B^\dagger$ form the bulk $SO(3)$ triplet
which corresponds to the edge $SO(3)$ triplet of
Majorana fermions discussed in [\ref{bib:fntw}]. The
fundamental charge $e$ {\it boson} is  created/annihilated by
${B^3}{B^\dagger}$, {\it i.\ e.\/} it is a dipole attached
to a flux tube.

Finally, in section \ref{sec:pair} we considered the alternative condensate
in which pairs of quasiparticles condensed in a singlet paired state.
We saw there that this condensate breaks $SU(2)_1 \times SU(2)_1 \to
SU(2)_2$. Hence, the order parameter of this condensate is in the
coset $SU(2) \times SU(2)/SU(2)$. However, this condensate does not have
topologically stable non-trivial soliton states since
$\pi_1(SU(2) \times SU(2)/SU(2))=0$. In contrast, the triplet condensate
discussed above does have topologically non-trivial solutions since
$\pi_1(SO(3) \times SO(3)/SO(3))={\bf Z}_2$, which carry the spin-$1/2$
representation. While both condensates break
$SU(2)_1 \times SU(2)_1 \to SU(2)_2$, the singlet paired condensate does
not support states with non-Abelian statistics while the triplet
(adjoint) condensate does. The crucial difference resides in the fact
that the triplet fields leave a ${\bf Z}_2$ subgroup of $SU(2)$
unbroken. The quasiparticles with non-Abelian statistics
carry the spin-$1/2$ representation which transforms non-trivially
under this ${\bf Z}_2$.  Therefore, the singlet condensates do not
have excitations  with non-Abelian statistics while the triplet
condensate does. From this point of view, the singlet condensates
(states with quasiparticle pairing) are actually Abelian FQH states.

\section{Effective edge theories}
\label{sec:edge}

\subsection{Edge theories for the $q$-Pfaffian}
\label{subsec:q}

The basic idea behind the Landau-Ginzburg theories presented
in the above has been that of a projection from an enveloping
Abelian theory. This mechanism can be beautifully illustrated
by considering effective edge conformal field theories (CFT)
for the various quantum Hall states that are involved.
In the conformal field theory setting, the notion
of obtaining a non-trivial theory by reduction from
a simpler one is an old idea known as the
 {\it conformal embedding}. For instance, the
${SU(2)_k}\times SU(k)_2$ current algebra -- which
does not have a free-field representation -- of the multi-channel
Kondo model can be embedded in an $SU(2k)_1$
current algebra -- which does. The projection
can be accomplished by standard techniques. In the preceding
sections, we have shown how the analogous projection
occurs in the bulk: it is not done `by hand', but
occurs dynamically as a result of spontaneous
symmetry-breaking. We found that this automatically
led to the physics of non-Abelian statistics. In this
section, we show how the same physics arises at the
edge in the context of a conformal embedding.

In such (conformal) edge theories,
the edge quasiparticles of non-Abelian quantum Hall states
exhibit `non-Abelian exclusion statistics'.
For the $q$-Pfaffian states at $\nu=1/q$, this has been
explained in a recent paper by one of us \cite{Sc2}.

What is important here is that the edge
theory for a Pfaffian quantum Hall state can be obtained by
a projection from a covering Abelian edge theory \cite{cappelli}.
The Abelian cover of the $q$-Pfaffian edge theory is the edge
theory for the Abelian bi-layer Halperin state $(q+1,q+1,q-1)$.
The latter state is
described by the inverse $K$-matrix
\begin{equation}
K^{-1} =
 {1 \over 4q} \left(\begin{array}{cc}
      1+q  & 1-q \\ 1-q & 1+q \end{array} \right)
\end{equation}
giving filling fraction
$\nu=\sum_{IJ}(K^{-1})_{IJ}=1/q$ and central charge $c=2$.

The mechanism by  which projecting a CFT down to a
smaller theory can induce non-Abelian statistics is by
now understood and well-documented \cite{cappelli1,cappelli2,BS}. The key
ingredient are the fusion rules for the CFT fields that
correspond to the quasiparticles of choice. For the case
of the $q$-Pfaffian, the half flux quantum quasihole (of charge
$+{e \over 2q}$) corresponds
to a conformal field of dimension $h={q+2 \over 16q}$.
In a formulation that employs the fields of a $c={1 \over 2}$
Ising CFT and a $c=1$ Gaussian CFT, the quasihole
field can be represented as
\begin{equation}
\psi_{\rm qh} =
 \sigma \, e^{ i {1 \over 2\sqrt{q}} \phi}
\label{pfqh}
\end{equation}
with $\sigma$ the (chiral) spin field of the Ising CFT
and $\phi(z)$ a chiral bosonic field.
In the edge CFT for the $(q+1,q+1,q-1)$ Halperin state,
the fundamental quasiholes have conformal
dimension $h={q+1 \over 8q}$ and are represented as
\begin{equation}
\psi_{\rm qh}^{(1,2)} = e^{ i l^{(1,2)}_J \phi^J}
\label{Abelianqh}
\end{equation}
with ${\bf l}^{(1)}=(1,0)$, ${\bf l}^{(2)}=(0,1)$,
and $\phi^J(z)$ two bosonic chiral fields.

The characteristic difference between the exclusion
statistics of the quasiholes in both cases follows by
inspecting their fusion rules. The Ising spin field
satisfies
\begin{equation}
[\sigma] \cdot [\sigma] = [{\bf 1}] + [\psi]
\label{fusion}
\end{equation}
with $\psi$ the Ising fermion
and the non-trivial right hand side leads to the non-Abelian
exclusion statistics of the quasiholes over the $q$-Pfaffian
\cite{Sc2}. The fact that
the right hand side of Eq.\ (\ref{fusion}) has two terms leads
to a degeneracy of $2^{n-1}$ for a state with $2n$ quasiholes
at fixed positions \cite{nw}. In thermodynamics this implies
that a single quasiholes carries an effective number of degrees
of freedom equal to $\sqrt{2}$.

In contrast, the fusion rules of the $(q+1,q+1,q-1)$
quasihole fields are additive in the ${\bf l}$ labels, and the
associated statistics are Abelian. In Haldane's terminology
\cite{Ha}, these quasiholes satisfy fractional exclusion
statistics with statistics matrix $G=K^{-1}$ \cite{FK}.

The reduction from the Halperin to the Pfaffian edge theory
is most beautifully explained by focusing on the special case
$q=1$. For this (bosonic) Pfaffian state, the edge theory is
identical to a $SU(2)_{k=2}$ (chiral) Wess-Zumino-Witten (WZW)
theory, the $SU(2)$
symmetry being identical to the `particle-hole' $SU(2)$ symmetry
described in section \ref{sec:1/2}. At the same time, the enveloping
$c=2$ CFT describes a $(2,2,0)$ state and is thus a product of
two copies of the $SU(2)_{k=1}$ chiral WZW theory (compare
with section \ref{sec:pair}).
It becomes clear then that the reduction from
the $(2,2,0)$ to the $q=1$ Pfaffian CFT is the complement
to the usual Goddard-Kent-Olive coset construction: one
rewrites the $SU(2)_1 \times SU(2)_1$ theory in terms of
$SU(2)_2 \times ({\rm Ising})$ and then projects out the
degrees of freedom corresponding to the Ising ($c={1 \over 2}$)
factor. This reduces the central charge from $c=2$ to
$c={3 \over 2}$ and it also reduces the pair of charge $e/2$
quasiholes Eq.\ (\ref{Abelianqh}) in the enveloping theory to the
single half flux quantum quasihole Eq.\ (\ref{pfqh}) in the
projected theory, in the process subtracting $1/16$ (which is
the dimension
of the Ising spin-field) from their conformal dimension.

For the general $q$-Pfaffian, there is no $SU(2)$ symmetry, but
the reduction process is essentially the same: the projection
amounts to splitting off an Ising CFT factor,  and it merges the
two elementary quasiholes over the Halperin state into a single one
in the projected Pfaffian theory, subtracting $1/16$ from their
conformal dimension.

For the charge $-e$  excitations at the edge one observes
a similar pattern. In the covering theory there are two such
excitations, with labels ${\bf l}=(q\pm1,q\mp1)$ and of conformal
dimension ${q+1 \over 2}$.
They project to the Pfaffian edge electron of
conformal dimension ${q+1 \over 2}$ (multiplying the identity
of the Ising factor) and to the $q$ flux quantum quasiparticles
of dimension ${q \over 2}$ (multiplying the Ising fermion).

Comparing with the bulk construction presented in the
above, we see a clear correspondence. In particular, the projection
onto $\sigma_x=+{1 \over 2}$ corresponds to the elimination
of the Ising degrees of freedom in the edge theory. The
statistical transmutation of one of the $SU(2)$ triplet fields
from bosonic to fermionic is at the edge effectuated by
splitting off and then discarding  the Majorana fermion
field of the Ising CFT, while the transmutation from Abelian
to non-Abelian of the quasihole statistics is at the edge
effectuated by splitting off an Ising spin field.

The bulk-edge correspondence established here is intuitively
quite appealing: the projection procedure which in the
bulk amounts to projecting an up-down `spin' degree of freedom
onto a definite direction, is at the edge effectuated by
eliminating an Ising CFT. In addition, the edge picture
clearly shows how a two-component quasihole with Abelian
statistics gets projected on a single quasihole which
represents an effective number of $\sqrt{2}$ degrees
of freedom and exhibits non-Abelian statistics. Finally,
the edge analysis reveals the physical reason for the
statistical transmutation of the triplet $(B^3,B^\dagger,B)$:
the elimination of an Ising CFT amounts to factoring out a
fermionic field and this changes statistics from
bosonic to fermionic.

\subsection{Extension to Read-Rezayi states}
\label{subsec:RR}

We observe that very similar
considerations apply to the order-$k$ non-Abelian
FQH Hall states recently proposed by Read and Rezayi
\cite{RR2}. These authors proposed a two-parameter family of
non-Abelian FQH states $\Psi_{M,k}$, with $M$ related to a
Laughlin exponent and $k$ the order of the `clustering'
that is allowed for fundamental fermions (bosons).
The filling factor for these states is $\nu={k \over Mk+2}$.
The $q$-Pfaffian states are the special case $k=2$,
$q=M+1$ of the Read-Rezayi states.

The edge theory for an order-$k$ Read-Rezayi state is obtained
by replacing the Majorana fermion in a Pfaffian edge theory
by a ${\bf Z}_k$ parafermion \cite{RR2}.
One can show that for the special case $M=0$, the edge CFT
for the Read-Rezayi state, at filling $\nu={k \over 2}$, becomes
identical
to an $SU(2)_k$ chiral WZW theory. Under this identification,
the ${1 \over k}$ flux quantum quasiholes over the Read-Rezayi
state correspond to the $j={1 \over 2}$ `spinon' excitations
of the $SU(2)_k$ WZW theory.

The non-Abelian exclusion statistics of quasiholes over
the Read-Rezayi FQH states can be studied by generalizing the
analysis of \cite{Sc2} for the Pfaffian ($k=2$).
For $M=0$ the quasihole statistics are the non-Abelian statistics
of the $j={1 \over 2}$ spinons in the $SU(2)_k$ WZW theory.
The latter have been analyzed in \cite{BLS,FS,BS},
the most important features being (i) a maximal occupation of
\begin{equation}
n^{\rm max} = 2k
\end{equation}
spinons per energy level, and (ii) an effective number of degrees
of freedom of
\begin{equation}
\alpha_k = 2\cos{\pi \over k+2}
\label{eq:alphak}
\end{equation}
per spinon. Both these features have a direct interpretation
in the quantum Hall context.

The maximal occupation $n^{\rm max}$
enters in the following expression for the Hall conductance
\begin{equation}
\sigma_H = n^{\rm max} \,{Q_{\rm qh}^2}\, {e^2 \over h}
\label{sigmaH}
\end{equation}
with $Q_{\rm qh}$ the quasihole charge. Putting in the value
$Q_{\rm qh}={1 \over 2}$ of the ${1 \over k}$ flux quantum quasiholes,
together with $n^{\rm max}=2k$, correctly reproduces the Hall
conductance $\sigma_H = {k \over 2}{e^2 \over h}$.
We stress that, as is illustrated by the Eq.\
(\ref{sigmaH}), it is the combination $(n^{\rm max}Q_{\rm qh}^2)$
rather than the quasiparticle charge $Q_{\rm qh}$ which enters
in the most basic phenomenology. For the case of
Abelian Laughlin states at $\nu=1/m$, this is
consistent with the {\it duality}\ \cite{nw2,es} between
charge $1/m$ quasiholes with $n^{\rm max}=m$ and charge
$-1$ electrons with $n^{\rm max}=1/m$, the two agreeing
on the value $(n^{\rm max} Q_{\rm qh}^2)$. A similar duality
has been demonstrated for the non-Abelian exclusion
statistics for the $q$-Pfaffian states \cite{Sc2}.
In a sense, the physical effects of fractional charge
can be traded for those of fractional statistics.

The quantity $\alpha_k$ of Eq.\ (\ref{eq:alphak})
is related to the number $d_N$ of excited states with $N=kn$
quasiholes at fixed positions. For $N$ large we have
$d_N \sim \alpha^N$. Clearly, $\alpha_k=1$ for the case
$k=1$, where the statistics are Abelian. A non-integer value
for $\alpha_k$, which expresses itself in thermodynamic
quantities such as 1-particle distribution functions,
is a hallmark feature of non-Abelian exclusion statistics.
A more detailed analysis leads to precise values for the
quantities $d_N$. For $k=2$ one recovers the value $2^{n-1}$
which we already cited in the above.
The prediction going with case $k=3$ ($d_{3n}=F_{3n-2}$
with $F_l$ the $l$-th Fibonacci number) has been
confirmed by a numerical analysis reported in \cite{RR2}.
We remark that while the value for $n^{\rm max}$ depends
on $M$, the degeneracy factor $\alpha_k$ is the same for
all $M$.

Clearly, the results of this paper, together with the simple
picture for the edge theories for the order-$k$ non-Abelian
FQH states, inspire an approach towards constructing effective
Landau-Ginzburg descriptions for these new quantum Hall
states. In section \ref{sec:pair} we indicated the first steps
for such a construction in the cases $M=0$. In the more
general cases with $M>0$ (which include the fermionic states),
there is no $SU(2)$ symmetry and the analysis is more involved.

\section{Conclusions}
\label{sec:conc}

In this paper, we have constructed a Landau-Ginzburg
theory for the universality class of
a Pfaffian state of bosons at $\nu=1$. This
construction elucidates the physics of this state
by giving a physical interpretation to
the $SU(2)$ symmetry,
casting light on the stability of the
universality class, emphasizing the pairing physics,
and clarifying the relation to an associated Abelian state.
In our concluding remarks, we would like to
dilate, albeit briefly, upon these issues.

The $SU(2)$ symmetry which was the linchpin
of the construction of \cite{fntw} descends
from the particle-hole symmetry of the Laughlin state
of bosons at $\nu=1/2$. This fact is
crystallized by our extraction of the
bosonic Pfaffian from the enveloping Abelian
state which has two copies of the bosonic
$\nu=1/2$ state and its appurtenant symmetries.
The $SU(2)$ symmetry, we emphasize, is not a symmetry
of the full spectrum of such a state --even if
the microscopic Hamiltonian is particle-hole
symmetric. Rather, it is a symmetry at the level
of the topological quantum numbers of the excitations.
In the bulk, for instance, this means that the
charged and neutral fermions have the same statistics
but their energies can be split. At the edge, this
means that the edge theory has an $SU(2)$ symmetry
which can be split by the velocities of the edge modes.
Nevertheless, the symmetry holds exactly for the
quantum numbers -- including and especially the
statistics -- in the bulk and at the edge.
For this reason, we expect that the non-Abelian
statistics which we have found will be impervious to
the vicissitudes of small symmetry-breaking
fields which may spoil the $SU(2)$ symmetry
at the energetic level. This is manifested by the
irrelevance of weak perturbations
of our final result, the $SU(2)_2$
Chern-Simons theory \cite{fntw}.

The bosonic Pfaffian state is a daughter
state of the $(2,2,0)$ state which results from
the condensation of a {\it neutral paired order
parameter}. As a result, the filling fraction
is unchanged. By contrast, the Abelian hierarchy
arises from condensation of a charged order
parameter. We have raised the possibility of
condensing another neutral order parameter
in the $(2,2,0)$ state which leads to a distinct
daughter state. One can contemplate
extending this notion to other parent states.
Suppose, for instance, that a neutral
quasiparticle-quasihole composite were
to condense in the $\nu=2/5$ state.
Would a non-Abelian state at the same
filling fraction result?

The condensation phenomena that we have discussed
results in the elimination of certain degrees of freedom,
thereby reducing an Abelian theory to a smaller non-Abelian theory.
This reduction is the bulk counterpart of
a corresponding edge construction, namely,
the projection from an enveloping Abelian
conformal field theory to a non-Abelian theory which is
conformally embedded in it.
In the case of the reduction from
the $(2,2,0)$ state to the Pfaffian, the
pseudo-spin degree of freedom is eliminated.
The Pfaffian state is the fully
polarized sibling of the unpolarized $(2,2,0)$
state. However, there is really a continuous
family of partially polarized states which interpolate
between them, as has been emphasized in \cite{ho,RR1},
and the state which corresponds to real experiments
could be somewhere in this family. Whether
this state is in the universality class
of the Pfaffian states or of their
enveloping -- or parent -- Abelian states
is a question which is answered by
the order parameter constructed
in this paper.

In some cases, such as the $\nu=5/2$ plateau,
there may not be such a pseudo-spin degree of freedom,
so the state may appear to be identically in
the fully polarized state. However,
one is tempted to speculate that a
layer structure could be spontaneously generated
by the dynamics in the $z$-direction (as
it is in wide quantum wells)
in a 2DEG so as to take advantage of
the energetics which favors the Pfaffian.

Of course, this entire construction
made heavy use of the $SU(2)$ symmetry of the
$\nu=1$ bosonic case. The extension to the
physically interesting case of fermions at $\nu=1/2$ is
still an open question to which we will return.

\section{Acknowledgements}

We thank Ed Rezayi for an early
communication of his work with Nick Read. EF and CN are
participants at the Program on {\it Disorder and Interactions
in Quantum Hall and Mesoscopic Systems} at the Institute for
Theoretical Physics of the University of California Santa
Barbara. This work was supported in part
by the NSF grant number NSF DMR94-24511 at UIUC (EF), and NSF
PHY94-07194 at ITP-UCSB (EF,CN). KS is supported in part by
the Foundation FOM of the Netherlands.

\end{multicols}
\end{document}